\documentclass[aps,prl,nobalancelastpage,superscriptaddress,twocolumn,longbibliography,nobibnotes,nofootinbib]{revtex4-1}

\usepackage[utf8]{inputenc}
\usepackage[american,british]{babel}
\usepackage[T1]{fontenc}
\usepackage[pdftex]{graphicx}  
\usepackage{xcolor}
\usepackage{dcolumn}
\usepackage{physics}
\usepackage{braket}
\usepackage{bm}
\usepackage{amsmath,amsthm,amssymb}
\usepackage{color}
\usepackage{verbatim}
\usepackage[normalem]{ulem}
\usepackage{dsfont}
\usepackage{hyperref}
\hypersetup{
 colorlinks=true,
 linkcolor=blue,
 anchorcolor = blue,
 citecolor = blue,
 filecolor = blue,
 urlcolor = blue
}

\def \be {\begin{equation}} 
\def \ee {\end{equation}} 
\def \l {\left(} 
\def \r {\right)} 
\def \la {\langle} 
\def \ra {\rangle}  

\date{}

\newcommand{\saclay}{Universit\'e Paris-Saclay, CNRS, LPTMS, 91405, Orsay, France.}

\begin{document}

\title{
Exceptional stationary state in a dephasing many-body open quantum system}
\date{\today}

\author{Alice March\'e}
\thanks{These authors contributed equally.}
\affiliation{\saclay}

\author{Gianluca Morettini}
\thanks{These authors contributed equally.}
\affiliation{\saclay}

\author{Leonardo Mazza}
\affiliation{\saclay}
\affiliation{Institut Universitaire de France, 75005 Paris, France.}

\author{Lorenzo Gotta}
\affiliation{Department of Quantum Matter Physics, University of Geneva, 1211 Geneva, Switzerland.}

\author{Luca Capizzi} 
\affiliation{\saclay}

\begin{abstract}
We study a dephasing many-body open quantum system that hosts, together with the infinite-temperature state, another additional stationary state, that is associated with a non-extensive strong symmetry. This state, that is a pure dark state, is exceptional in that it retains memory of the initial condition, whereas any orthogonal state evolves towards the infinite-temperature state erasing any information on the initial state.
We discuss the approach to stationarity of the model focusing in particular on the fate of interfaces between the two states. 
A simple model based on a membrane picture helps developing an effective large-scale theory, which is different from the usual hydrodynamics since no extensive conserved quantities are present.
The fact that the model reaches stationary properties on timescales that diverge with the system size, while the Lindbladian gap is finite, is duly highlighted.
We point out the reasons for considering these exceptional stationary states as quantum many-body scars in the open system framework.
\end{abstract}

\maketitle 

\paragraph{\textbf{Introduction ---}}

The eigenstate thermalization hypothesis (ETH) stands as a cornerstone in condensed-matter physics, as it describes in simple and physical terms the emergence of thermal features in the late-time dynamics of closed quantum systems~\cite{berry1977, Deutsch-91, Srednicki-99, D_Alessio_2016, pfk-22}.
In the last years, exceptional dynamics that escape thermalization by retaining memory of the initial conditions, have been identified as a primary mechanism for violating the ETH, via rare eigenstates dubbed many-body quantum scars~\cite{Bernier-17, Turner_2018, mm-20, Serbyn-21, Moudgalya_2022, Chandran-23}.
The broad goal of this article is to study a similar phenomenon in Markovian many-body open quantum systems (MBOQS)~\cite{fazio2024manybodyopen,sbmd-23}, discussing a specific dephasing model.  
 
The late-time behaviour of Markovian MBOQS~~\cite{Kessler_2012, Minganti_2018}, has recently attracted a great deal of attention~\cite{Poletti_2012, Poletti_2013, cb-13, Bouganne_2019, Wellnitz_2022, LiSalaPollmann_2023, Brighi2024,li2024highlyentangled}.    
In the absence of any strong symmetry~\cite{Buca_2012, Albert_2014}, one generically expects a unique stationary state that is progressively approached from any initial state. This leads to the complete erasure of any information on the initial state, a behaviour that resembles closely the ergodicity that always accompanies thermalization in closed systems.
Whether a MBOQS  could also withstand an exceptional stationary state is something that has not been discussed yet.

In this article we study a one-dimensional dephasing model where  the infinite-temperature (mixed) state and the fully-polarized (pure) state are the only stationary states.
The latter is a dark state of the model, but we argue that additionally it plays the role of a scar in this MBOQS because it is not protected by \textit{extensive} strong symmetries~\cite{mm-23a}. 
In fact, it is protected by a \textit{non-extensive} symmetry: no spatial density of conserved quantity exists that could give rise to transport phenomena.
Given that we can also demonstrate the presence of a finite spectral gap of the Lindbladian, these considerations rule out the presence of hydrodynamic slow modes.
Nonetheless, we show that the interface between the two stationary states diffuses according to the physics of a fluctuating membrane,
%We show that, despite the lack of conservation laws, usually associated with the hydrodynamics of extensive conserved charges, diffusive scalings are observed and their origin is traced back to a fluctuating membrane. 
%Surprisingly, such
in a scenario characterized by long relaxation times, 
and we show how it is consistent with the finite Lindbladian gap (see also~\cite{Znidaric-15,ms-20, Haga_2021, Rakovszky_2024}).

\begin{comment}
The relaxation time of local observables can be algebraically long in the system size, albeit a finite spectral gap in the Lindbladian is present; a similar observation has been pointed out in different contexts in Refs.~\cite{Znidaric-15,ms-20, Haga_2021, Rakovszky_2024}.
To understand the mechanism underlying the dynamics, we develop a membrane picture that describes the fate of interfaces between the two steady states: their ballistic drift and diffusive fluctuations are uncovered, showcasing a hydrodynamic behaviour that is not associated to any local conserved quantity.
Our work proposes a novel view on many-body quantum scars in MBOQS as \textit{exceptional stationary states} that parallels the weak ETH violation associated to scars in the unitary context. 
\end{comment}

\paragraph{\textbf{The model ---}}

We consider a one-dimensional spin-1/2 chain of length $L$ whose density matrix $\rho$ evolves according to the Lindblad master equation
\begin{equation}
\label{eq:model}
\frac{d}{dt}\rho = \mathcal{L}[\rho] = -i[H,\rho] + \sum_j L_j \rho L^\dagger_j - \frac{1}{2}\{L_j^\dagger L_j,\rho\},
\end{equation}
with jump operators $L_j = \sqrt{\gamma} \sigma^z_j$ and Hamiltonian
\begin{equation}
\label{eq:H}
 H =  \sum_j 
J\left[
\sigma^{+}_j \sigma^{-}_{j+1}+H.c.
\right]+
 g \, \left[ \sigma^x_j \pi^z_{j+1}
 + \pi^z_j \sigma^x_{j+1}  \right],
\end{equation}
where $\pi_j^{\alpha} = \mathds{1}- \sigma^\alpha_j$, for $\alpha = x,y,z$.
Here, $J$ is the strength of the nearest-neighbor hopping term, while $g$ multiplies the East-West Hamiltonian~\cite{Pancotti-20}, according to which a spin can be flipped only when a neighboring spin is in a $\ket{\downarrow}$ state. 
The dissipation occurs at a rate $\gamma$ and it is associated with the decay of coherence in the $z$-polarized basis; as such it is naturally interpreted as a \textit{dephasing} process. For $g=0$ the model can be solved analytically with Bethe-ansatz techniques \cite{mep-16}, and, unless otherwise specified, we focus here on $g,J\neq 0$.

\paragraph{\textbf{Stationary states} ---}

\begin{figure}[t]
 \includegraphics[width=\columnwidth]{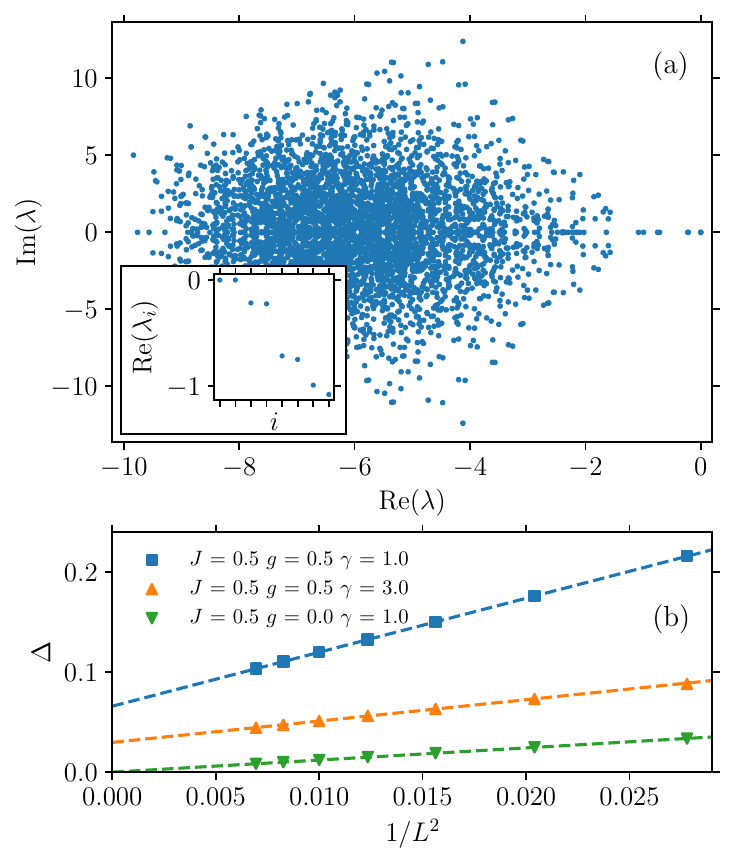}
 \caption{(a) Lindbladian spectrum for the following parameters: $[J,g,\gamma,L]=[0.5,0.5,1,6]$. Inset: real parts of the eigenvalues closest to zero, labeled by the index $i$.
(b) Lindbladian gap $\Delta$ as a function of $1/L^2$ for $L \in \{ 6,\ldots,12 \}$. The results are compatible with a scaling $\Delta(L) = \Delta_0 + \Delta_1 / L^2 + o(L^{-2})$. The three curves correspond to different sets of parameters: $[J,g,\gamma]=[0.5,0.5,1]$, $[0.5,0.5,3]$ or $[0.5,0,1]$.
For both panels (a) and (b), open boundary conditions are considered.
}
\label{fig:liouvillian_gap}
\end{figure}

The infinite temperature state $\rho_{\infty} = \mathds{1}/\text{Tr}[\mathds{1}]$ is a stationary state of~\eqref{eq:model}, as a consequence of $L_j = L_j^\dagger$. Furthermore, given the specific choice of the Hamiltonian and dissipation the pure state $\rho_\Uparrow = \ketbra{\Uparrow}{\Uparrow}$ associated with all spins aligned along the $z$ axis, is also stationary and is a dark state of the model~\cite{Kraus-08}. 
Note that the two states are not orthogonal according to the Hilbert-Schmidt scalar product $\langle \rho, \tau \rangle_{\rm HS} := \text{Tr}[\rho^\dagger \, \tau]$ because $\langle \rho_{\Uparrow} , \rho_\infty \rangle_{\rm HS} = 2^{-L} $;
hence it is convenient to consider $\rho'_\infty := (\mathds{1} - \rho_\Uparrow ) / \text{Tr}[\mathds{1} - \rho_\Uparrow]$, that is orthogonal to $\rho_\Uparrow$ but whose local properties differ from those of $\rho_\infty$ by a term that is exponentially small in the system size. 

We point out that $\rho_{\Uparrow}$ is technically a strong symmetry (i.e.~a conserved quantity of the MBOQS~\cite{Buca_2012, Albert_2014}) of the model, meaning that it commutes with both $H$ and $L_j$ $\forall j$; nonetheless, 
since it cannot be written as $\sum_j q_j$, with $q_j$ a local operator, it is non-extensive.
When $g=0$, instead, the total magnetization is an \textit{extensive} strong symmetry.

\begin{comment}
Specifically, as it is always true when $L_j=L^{\dagger}_j$, the \textit{infinite temperature state} $\rho_{\infty} = \mathds{1}/\text{Tr}[\mathds{1}]$ is stationary, that is $\mathcal{L}[\rho_\infty]=0$. 
Instead, the magnetisation $\sum_j \sigma_j^z$ is not, in general, a strong symmetry and it evolves non-trivially during the open dynamics. Nonetheless, given the specific choice of Hamiltonian~\cite{Pancotti-20, Bocini_2024} and dissipation in Eq.~\eqref{eq:H}, the pure state $\rho_\Uparrow = \ketbra{\Uparrow}{\Uparrow}$ associated with all spins aligned along the $z$ axis, is stationary, as it constitutes a non-trivial strong symmetry of the model; in particular, the subspace generated by $\ket{\Uparrow}$ is decoherence-free~\cite{Lidar-12}.
Note that the two states are not orthogonal according to the Hilbert-Schmidt scalar product $\langle \rho, \tau \rangle_{\rm HS} := \text{Tr}[\rho^\dagger \, \tau]$ because $\langle \rho_{\Uparrow} , \rho_\infty \rangle_{\rm HS} = 2^{-L} $;
hence it will be more useful in the following to consider $\rho'_\infty := (\mathds{1} - \rho_\Uparrow ) / \text{Tr}[\mathds{1} - \rho_\Uparrow]$, whose local properties differ from those of $\rho_\infty$ by a term that is exponentially small in the system size but that is orthogonal to $\rho_\Uparrow$.
\end{comment}

No additional states are stationary. To prove that, we consider the Lindbladian $\mathcal L$, which is a linear non-Hermitian superoperator, 
and study its complex spectrum, plotted in Fig.~\ref{fig:liouvillian_gap}(a), obtained with the exact-diagonalization (ED) of a spin chain of size $L=6$. The calculation is performed with the open-source python-framework QuTiP~\cite{Qutip1, Qutip2}.
The inset shows that the eigenvalue $\lambda = 0$ is two-fold degenerate, 
and we have checked that the two associated eigenvectors can be identified with $\rho_{\infty}$ and $\rho_{\Uparrow}$. This feature is compatible with properties of the commutant algebra~\cite{mm-22, mm-23, mm-23a, LiSalaPollmann_2023} of the model, derived in the End Matter (EM) and detailed in the Supplemental Material (SM)~\cite{ref_SM}.
The numerics allows us also to conclude that the Lindbladian spectral gap, $\Delta = \min_{\lambda \neq 0} |\Re[\lambda]|$ does not close with system size, and specifically that it has a large-$L$ scaling $\Delta(L) \simeq \Delta_0 + \frac{\Delta_1}{L^2}$ with $\Delta_0 \neq 0$ whenever $g \neq 0$~\footnote{We remark that, at $g=0$ the degeneracy of the eigenvalue $\lambda=0$ is enhanced, due to the $U(1)$ symmetry associated with the conservation of the magnetization along the $z$-axis}.   
The numerical scaling of $\Delta(L)$ is presented in Fig.~\ref{fig:liouvillian_gap}(b) for sizes up to $L=12$.
The specific value of $\Delta_0$ depends on the choice of open versus periodic boundary conditions, see EM.

\paragraph{\textbf{Dynamics} ---}

We argue that any initial pure state $\ket{\psi}$ that is orthogonal to $\ket{\Uparrow}$ must relax eventually towards $\rho_\infty'$. The approach to $\rho_\infty'$ can be quantified by the normalized Hilbert-Schmidt scalar product:
\begin{equation}
 \theta_{\infty}' (t) := \frac{\text{Tr}[\rho'_{\infty} \rho(t)]}
 {\sqrt{\text{Tr}[\rho^{\prime 2}_{\infty}] \text{Tr}[\rho(t)^2]}},
\end{equation}
that generalizes the scalar product $|\langle \psi | \psi' \rangle|^2$ (which holds for pure states).
Since $\rho_\infty'$ and $\rho(t)$ are positive and Hermitian operators, $\theta_\infty'(t) \geq 0$.
Moreover, by the Cauchy-Schwartz inequality $ \theta'_\infty(t) \leq 1$ and
$\theta'_{\infty}(t) = 1$ if and only if $\rho(t) \equiv \rho'_{\infty}$; in particular, whenever $\bra{\psi} \Uparrow \rangle = 0$, $\theta'_\infty(t)$ is expected to converge to $1$ at sufficiently large time $t$.

We employ the ITensor julia library~\cite{itensor, itensor-r0.3} and develop a tensor-network representation of the density matrices $\rho'_{\infty}$ and $\rho(t)$, where the latter is computed with an algorithm based on the time-dependent variational principle~\cite{TDVP2_2020, Haegeman2011} that integrates the dynamics~\eqref{eq:model}; see EM for details. The scalar product $\theta_{\infty}'(t)$ can be efficiently computed and we can study chains of length $L=40$.

\begin{figure}[t]
 \includegraphics[width=\columnwidth]{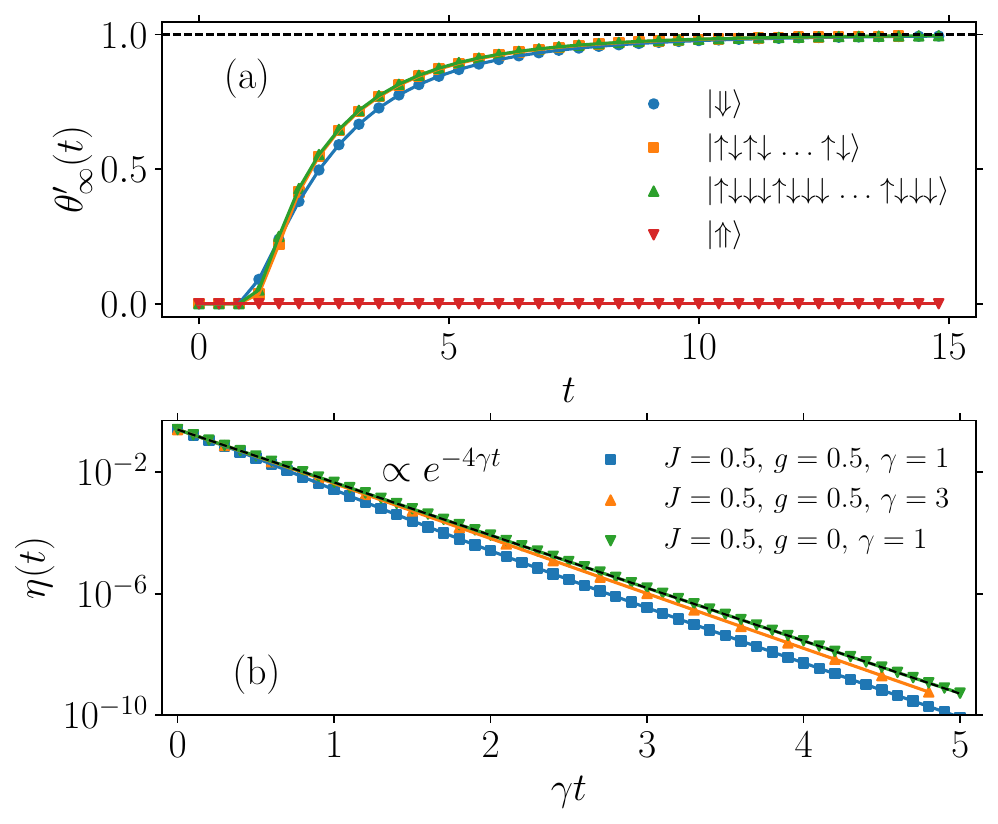}
     \caption{(a) Time-evolution of $\theta'_\infty(t)$ for the initial states $\ket{\Downarrow}$, $\ket{\text{N\'eel}}$, $\ket{\uparrow \downarrow \downarrow \downarrow  \uparrow \downarrow \downarrow \downarrow \ldots}$, and $\ket{\Uparrow}$. The parameters are $[J,g,\gamma,L]=[0.5,0.5,1,40]$. (b) Time-evolution of $\eta(t)$ for the initial state $\ket{\uparrow \ldots \uparrow \rightarrow \uparrow  \ldots \uparrow}$. We observe that the upper bound is saturated for $g=0$.}
 \label{Fig:2}
\end{figure}

Our results are plotted in Fig.~\ref{Fig:2}(a) and demonstrate the time-evolution towards $\rho'_{\infty}$ of three 
%translationally-invariant 
initial states orthogonal to $\ket{\Uparrow}$, namely the fully-polarized state $\ket{\Downarrow}$ with all spins in the $\ket{\downarrow}$ state, the antiferromagnetic N\'eel state $\ket{\text{N\'eel}}=\ket{ \uparrow \downarrow \uparrow \downarrow \uparrow \downarrow \ldots }$, and the state $\ket{\uparrow \downarrow \downarrow \downarrow  \uparrow \downarrow \downarrow \downarrow \ldots}$. We also verified numerically the fact that $\ket{\Uparrow}$ is stationary and retains $\theta_\infty'(t)=0$ at all times.

These results highlight the separation of the Hilbert space into two disconnected subspaces. 
On the one hand, we have the state $\ket{\Uparrow}$ and the associated subspace $\mathcal S_{\Uparrow}$. 
On the other hand, the subspace $\mathcal T_\Uparrow$, orthogonal to $\mathcal S_{\Uparrow}$, is characterized by a loss of information on the initial conditions, and an eventual relaxation towards $\rho'_{\infty}$.

We now show that any coherence between the two subspaces $\mathcal T_{\Uparrow}$ and $\mathcal S_{\Uparrow}$ is lost exponentially in time by introducing the quantifier:
\begin{equation}\label{eq:eta_t}
 \eta(t) :=
 \sum_{\ket{\boldsymbol \sigma} \neq \ket{\Uparrow}}
 \left| 
 \bra{\boldsymbol \sigma} \rho(t) \ket{\Uparrow}
 \right|^2,
\end{equation}
where $\{ \ket{\boldsymbol \sigma} \}$ is a basis for the Hilbert space of the entire spin chain composed of states that are simultaneous eigenstates of all $\sigma^z_j$;
the sum is restricted over a basis of $\mathcal T_{\Uparrow}$.
With a few simple algebraic passages, we obtain the following suggestive rewriting: $\eta(t) = \bra{\Uparrow} \rho(t)^2 \ket{\Uparrow} - 
\bra{\Uparrow} \rho(t) \ket{\Uparrow}^2$, which proposes the simple interpretation of $\eta(t)$ as the variance of the operator $\rho(t)$ on the state $\ket{\Uparrow}$, and proves that the definition of $\eta(t)$ is basis independent. 
By definition, $0 \leq \eta(t) \leq 1$.

In the EM, we detail a few algebraic passages that demonstrate the inequality 
$ \frac {d}{dt} \eta(t) \leq - 4 \gamma \eta (t)$,
and thus that $\eta(t) \leq \eta(0)e^{- 4 \gamma t}$, proving an exponential decay in time of the coherence between the two subspaces associated with the two different dynamics.
We explicitly verified this prediction on a system initialized in the state $\ket{\uparrow \ldots \uparrow \rightarrow \uparrow  \ldots \uparrow}$, $\rightarrow$ denoting the spin aligned along the $x$-axis, see Fig.~\ref{Fig:2}(b).

\paragraph{\textbf{Open-system quantum many-body scars} ---}

The results presented so far motivate the interpretation of $\ket{\Uparrow}$ as an \textit{open-system generalization of isolated quantum many-body scars}~\cite{Pancotti-20, Bocini_2024}, as we are now going to explain. 

We first remark that, in general, many-body systems can possess extensive conserved quantities generated by local densities, $Q_a= \sum_j q_{a,j}$. 
For closed systems, the (generalized) Gibbs ensembles $\la \dots\ra \propto \text{Tr}[e^{-\sum_a \beta_a Q_a}\dots]$ are the stationary states determined the $Q_a$s. 
They are crucial for the appearance of a standard hydrodynamic behaviour, 
based on the description of the system in terms of local Gibbs ensembles~\cite{km-63,Spohn-12,Doyon-20}, with $\beta_a(x,t)$ depending smoothly on coarse-grained spatial and time variables. 
According to strong ETH, these are the only possible stationary states~\cite{Srednicki-99, D_Alessio_2016}; in particular, when only $H$ is conserved, the eigenstates of the Hamiltonian are expected to be locally indistinguishable from Gibbs ensembles at the corresponding energy density. 
Violations of such a scenario are possible, for instance, via the presence of eigenstates whose local properties are not described by Gibbs ensembles, and are dubbed \textit{quantum many-body scars}~\cite{Serbyn-21, Moudgalya_2022, Chandran-23}. 
These states are associated to non-extensive symmetries~\cite{mm-23a}.

Following this analogy, we propose to identify scars in MBOQS 
characterized by dephasing jump operators
as \textit{stationary states that are not protected by extensive conserved operators}. 
We have already seen that when $g\neq 0$, $\ketbra{\Uparrow}{\Uparrow}$ is a conserved operator (strong-symmetry) which is not extensive~\cite{Bocini_2024}:  we identify $\ket{\Uparrow}$ as an \textit{open system quantum many-body scar}. 
In contrast, for $g=0$ the conservation of the magnetization $S^z$, which is extensive, protects $\ket{\Uparrow}$, and we refrain from calling it scar: it would just be a dark state. 
Moreover, the absence of extensive conserved operators rules out the existence of a standard late-time hydrodynamics and of its slow modes;
this is compatible with the accepted expectation that the finite Lindbladian spectral gap plotted in Fig.~\ref{fig:liouvillian_gap}(b) should force a relaxation on a finite timescale in the thermodynamic limit (note instead that for $g=0$ the gap is zero, compatibly with the hydrodynamics of $S^z$).
% We also stress that the validity of the usual hydrodynamics of extensive conserved charges, which are indeed absent in our model, is essentially ruled out. 

Finally, we point out that a rigorous and widely accepted definition of scars in MBOQS is, up to our knowledge, absent, and different points of view on weak ergodicity breaking in open quantum systems have been proposed in Refs.~\cite{Wang_2024,deLeeuw_2024,Fernandez-Hurtado_2014, Bu_a_2019, Qianqian_2023}.

\paragraph{\textbf{Membrane diffusion of the interface} ---}

\begin{figure}[t]
 \includegraphics[width=\columnwidth]{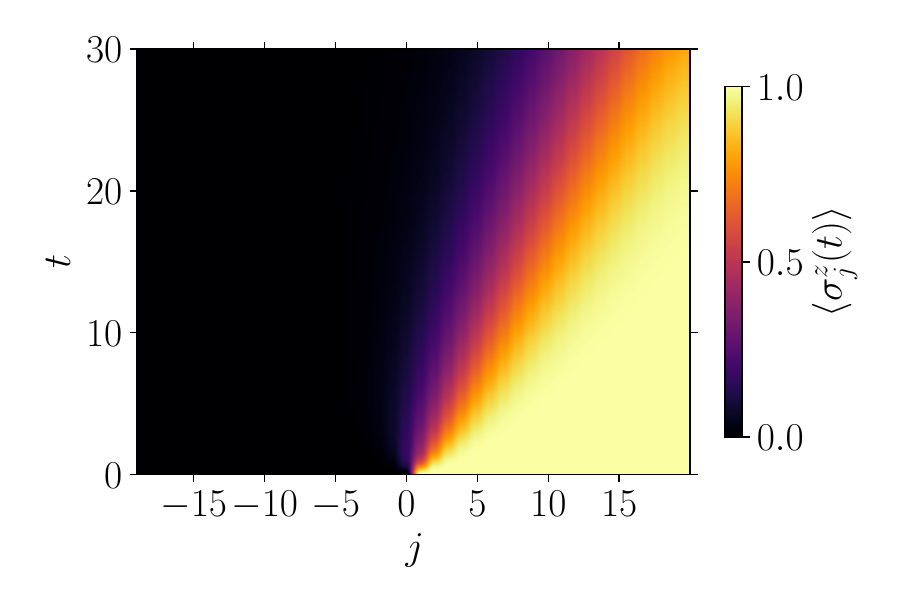}
 \caption{Magnetization profile $\langle \sigma^z_j (t)\rangle$ as a function of $j$ and $t$. The initial state is obtained by joining the infinite-temperature state, with support $j \in [-L/2+1,0]$, and the full-aligned up state along the $z$-axis elsewhere. The evolution is performed for the model in Eqs.~\eqref{eq:model} with the following parameters: $[J,g,\gamma,L]=[0.5,0.5,1,40]$. }
 \label{fig:xt_profile}
\end{figure}

We now discuss the existence of slow decay modes also in model~\eqref{eq:model}, notwithstanding the finite spectral gap and the absence of hydrodynamics.
%To understand the interplay between the two stationary states of the model, w
We consider a bipartition protocol where we juxtapose the infinite temperature state and the scar (on the left/right half-chain respectively):  with a slight abuse of notation, we denote the initial state as $\rho(0) = \rho_\infty \otimes \rho_\Uparrow$. 
%This allows understanding of both the thermalization process in the presence of an isolated scar and the large-scale dynamics of the system. 
We show the numerical results for the local magnetization $\langle \sigma^z_j (t)\rangle := \text{Tr} [\sigma_j^z \rho(t)]$ in Fig.~\ref{fig:xt_profile}, displaying a progressive melting of the scar, as expected for the eventual thermalization. 
However, the picture suggests that thermalization is slowly achieved on a time-scale diverging as $L$; while the diffusive broadening resembles standard hydrodynamics, we have ruled out such scenario. 
We instead propose a \textit{membrane picture}~\cite{nrvh-17,zn-20}
to describe the fluctuating dynamics of the interface separating the two stationary states.

We are able to pinpoint analytically this picture in the large-$\gamma$ limit of the dynamics~\eqref{eq:model} using the second-order perturbation theory for $g/\gamma$, $J/\gamma\ll 1$ introduced in Ref.~\cite{cb-13}  for dephasing models.
It is based on an effective Lindbladian $\mathcal{L}_{\text{eff}}$ that acts onto 
the states $\rho_{\boldsymbol{\sigma}}=\ketbra{\boldsymbol{\sigma}}{\boldsymbol{\sigma}}$, which are easily proven to be the only stationary states for $g,J=0$ and thus determine the late-time dynamics $t \gg \gamma^{-1}$ whenever $g/\gamma$, $J/\gamma\ll 1$.

The action of $\mathcal L_{\text{eff}}$ onto a 
density matrix
$
\rho(t) = \sum_{\boldsymbol{\sigma}}p_{\boldsymbol{\sigma}}(t)\ketbra{\boldsymbol{\sigma}}{\boldsymbol{\sigma}},
$
is most easily described using the Doi-Peliti formalism that represents the density matrix $\rho(t)$ as a vector $| \rho(t) ) =\sum_{\boldsymbol{\sigma}}p_{\boldsymbol{\sigma}}(t) |\boldsymbol{\sigma})$ that satisfies the unconventional normalization condition $\sum_{\boldsymbol{\sigma}} p_{\boldsymbol \sigma}=1$~\cite{Doi-76,Doi-76a}. 
The Lindblad dynamics $\frac d{dt} \rho(t) = \mathcal L_{\text{eff}}[\rho(t)]$ is translated into the form $\frac{d}{dt}|\rho(t)) = -W |\rho(t))$
with the Markov generator (see EM and SM~\cite{ref_SM})
\begin{align}\label{eq:W}
W =& -\frac{J^2}{4\gamma} \sum_j \left(\sigma^x_j\sigma^x_{j+1}+\sigma^y_j\sigma^y_{j+1}+\sigma^z_j\sigma^z_{j+1}-1 \right)+ \nonumber \\
&+2\frac{g^2}{\gamma} \sum_j \left( \pi^{x}_j \pi^z_{j+1}ù + \pi^{x}_{j+1} \pi^z_{j}+ \pi^z_{j-1}\pi^x_{j} \pi^z_{j+1}\right).
\end{align}

Since the dynamics induced by $W$ cannot be determined analytically, we propose a simplified but analytically-tractable generator $\tilde W = W- 2 (g^2 / \gamma) \sum_j \pi^z_{j-1} \pi^x_j \pi^z_{j+1}$ 
without terms coupling three sites.
We also make a specific choice of the parameters, namely
$J^2/4 \gamma = 1/6$ and $2 g^2 / \gamma = 1/6$: this allows us to provide rigorous estimates for the spectral gap (see SM~\cite{ref_SM}) and exact calculations for the interface dynamics.

We introduce the vector $|\bullet) = \frac{1}{2}\l |\uparrow ) + | \downarrow )\r$, namely the local infinite temperature state, and we define the state
$ |x) \equiv  
| \bullet\dots \bullet \underset{x}{\bullet} \uparrow \uparrow \dots \uparrow )$,
representing an interface between the positions $x$ and $x+1$. 
The states $|x)$ are the essence of the membrane picture. 
By explicit computation,
$
\tilde{W} |x) = |x)
-\frac{2}{3}|x+1)-\frac{1}{3}|x-1),
$
and therefore the space of single domain-wall states is closed under Markov dynamics; in particular, the evolution reduces to a one-dimensional Brownian motion with drift on the lattice, which can be solved exactly using the Fourier transform. The explicit solution is $|\rho(t)) = e^{-\tilde{W}t}|x=0) = \sum_x p(x,t)|x)$
with $
% \label{eq:1kink_solution}
p(x,t)\equiv \int^{\pi}_{-\pi}\frac{dk}{2\pi}\exp\l -\varepsilon(k)t+ikx\r,
$
where 
$
\varepsilon(k) = 1-\cos(k)+\frac{i}{3}\sin k.
$
Here, $p(x,t)$ is the probability distribution of the domain-wall position; in the large $t$ limit, we approximate it as
\be
p(x,t) \simeq \frac{1}{\sqrt{2\pi \mathcal{D}t}}\exp\l -\frac{(x-vt)^2}{2\mathcal{D}t}\r.
\ee
where $v=1/3$ and $\mathcal{D}=1$, the \textit{drift velocity}  and the \textit{diffusion constant}  respectively, are defined by the small $k$ expansion
$
\varepsilon(k) \simeq ivk +\frac{1}{2}\mathcal{D}k^2 + O(k^3)
$.

%We now discuss the time evolution of the expectation value of $\sigma^z_j$ plotted in Fig.~\ref{fig:xt_profile}.
%which is a local classical observable (namely, it is diagonal in the $\{ |\boldsymbol \sigma) \}$ basis).
% We call a \textit{classical observable} one that is diagonal in the $\{ |\boldsymbol \sigma) \}$ basis, and thus that reads
% $\mathcal{O} \equiv \sum_{\boldsymbol{\sigma}} \mathcal{O}_{\boldsymbol{\sigma}} |\boldsymbol{\sigma})(\boldsymbol{\sigma}|$: it has expectation value equal to 
% $\langle \mathcal {O} \rangle = \sum_{\boldsymbol{\sigma}}p_{\boldsymbol{\sigma}}\mathcal{O}_{\boldsymbol{\sigma}}$.
The expectation value of the local magnetization operator $\sigma^z_j$ plotted in Fig.~\ref{fig:xt_profile} has a simple representation using membrane states,
$ \langle \sigma^z_j(t)\rangle =  \sum_{j' <j}p(j',t)$;
in the large-scale coarse-grained limit where the discrete lattice site $j$ is replaced by a continuous variable $x$, it can be approximated by the Error function:
\begin{equation}
\langle \sigma^z(x,t) \rangle \simeq 
\int^{\frac{x-vt}{\sqrt{\mathcal{D}t}}}_{-\infty}\frac{du}{\sqrt{2\pi}} \ \exp\l -\frac{u^2}{2}\r.
\label{Eq:erf:rescaling}
\end{equation}
In general, at late times, we can interpret the local state across the interface as a statistical mixture of the two stationary states.

%%%%FIG
\begin{figure}[t]
 \includegraphics[width=\columnwidth]{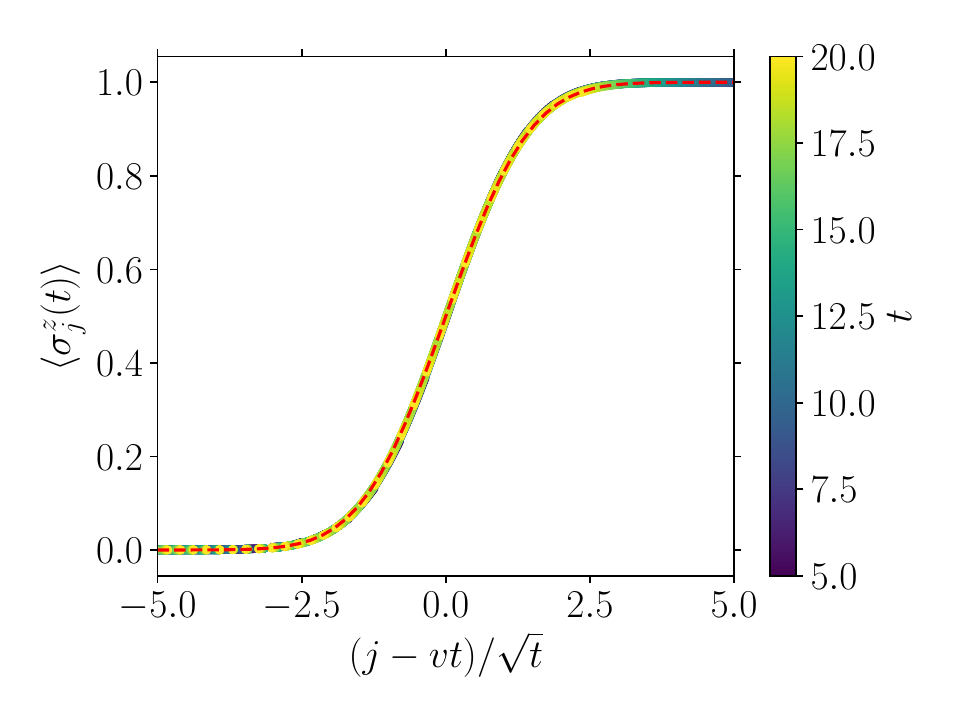}
 \caption{Profile of the magnetization shown in Fig.~\ref{fig:xt_profile} as a function of the scaling variable $(j-vt)/\sqrt{t}$. The rescaled data collapse onto the universal function described by Eq.~\eqref{Eq:erf:rescaling}, represented by the red dashed line. The velocity $v$ is determined by interpolating numerically, at each time $t$, the position $x$ where the profile satisfies $\langle \sigma^z_j(t)\rangle = 0.5$. The diffusion constant $\mathcal{D}$ is a fit parameter. Here, $v \simeq 0.453$ and $\mathcal{D} \simeq 1.33$. Parameters used: $[J,g,\gamma,L]=[0.5,0.5,1,40]$.}
 \label{fig:rescaling}
\end{figure}
%%%%%

In Fig.~\ref{fig:rescaling} we show that the membrane picture gives an accurate description of the dynamics and show that the interface is actually evolving via a diffusive broadening with drift, even if we are not in the large-dissipation limit of the analytical picture and we simulate the full Lindbladian.
The collapse of the data presented in Fig.~\ref{fig:xt_profile} against the rescaled variable $(x- vt) / \sqrt{\mathcal{D} t}$, where $v$ and $\mathcal{D}$ have been fitted, is excellent. In order to show that this behavior is not a peculiar feature of our model, in the SM~\cite{ref_SM} we present numerical simulations for other two models.

\paragraph{\textbf{Late-time asymptotics and spectral gap} ---}

We conclude by reconcyling the slow dynamics we just computed with the presence of a finite Lindbladian gap  $\Delta_0$, following a reasoning that is reminiscent of Ref.~\cite{Haga_2021}.
In general, $\Delta_0$ is interpreted as a finite asymptotic decoherence rate~\cite{Kessler_2012, Minganti_2018}, or, equivalently, as a typical time-scale $\tau_0 \sim 1/\Delta_0$ that does not depend on the system size and that gives an upper bound to the time that is necessary for the system to reach stationarity. 
On the contrary,
our numerics in Fig.~\ref{fig:xt_profile} and the membrane analytical picture show that the thermalization of the interface requires a timescale that necessarily grows with the size of the system ($v$ and $\mathcal D$ are finite in the thermodynamic limit).
This suggests that the identification of the \textit{bona fide} asymptotic decay rate in a MBOQS requires a careful study not only of the Lindbladian spectral properties, but also of the size-dependence of the decomposition of initial states onto the eigenvectors of the Lindbladian~\cite{Znidaric-15,ms-20, Haga_2021, Rakovszky_2024}.

To do so, we investigate the finite-size behaviour of the interface dynamics; specifically, we focus on the spectral properties of the Markov operator $\tilde{W}$ with open boundary conditions. We express the evolution of a local observable $\mathcal{O}$, with $|x_0)$ the initial interface state, as
\be\label{eq:diff_ss}
\la\mathcal{O}(t)\ra - \la\mathcal{O}(t=\infty)\ra = \sum_{n\neq 0} e^{-\epsilon_n t}(\mathds{1}|\mathcal{O}|n)(\tilde{n}|x_0).
\ee 
Here, $\epsilon_n$ are the eigenvalues of $\tilde{W}$, $\{|n)\}_n,\{(\tilde{n}|\}_n$ its right/left eigenvectors respectively~\cite{Ashida_2020} and $(\mathds{1}| := \sum_{\boldsymbol{\sigma}} (\boldsymbol{\sigma}|$. 
We checked that the matrix $\tilde W$ has a real spectrum even if it is non-symmetric, and its nonzero eigenvalues are separated by a finite gap $\Delta$. 
As a consequence, we estimate an eventual exponential decay of \eqref{eq:diff_ss} via
\be\label{eq:diff_ss1}
|\la\mathcal{O}(t)\ra - \la\mathcal{O}(t=\infty)\ra| \leq e^{-\Delta t} \sum_{n\neq 0} |(\mathds{1}|\mathcal{O}|n)(\tilde{n}|x_0)|.
\ee
However, the magnitude of the matrix elements appearing above grows exponentially in the system size as $|(\mathds{1}|\mathcal{O}|n)(\tilde{n}|x_0)| \sim e^{O(L)}$ 
in a phenomenology related to that of the \textit{non-Hermitian skin effect}~\cite{gohsrich2024nonhermitianskineffectperspective}
(details are in the EM). As a consequence, the bound \eqref{eq:diff_ss1} can only be useful for time scales $t \gtrsim L$ and is otherwise too loose. This explains the apparent discrepancy between the spectral gap and the relaxation time of local observables.

\paragraph{\textbf{Conclusions} ---}

We have discussed a MBOQS dephasing model with an exceptional stationary state $\ket{\Uparrow}$ that retains the memory of the initial condition while any other initial state relaxes towards the infinite-temperature state $\rho'_\infty$.
We have employed this model to stress a viewpoint on quantum many-body scars in open systems based on the absence of extensive strong symmetries and of the associated long-time hydrodynamics.
Nonetheless, the scar is proven responsible for a slow-relaxation phenomenon that is based on the membrane diffusion mechanism. The work opens the path toward the study of dynamics in the presence of exceptional stationary states, the universal features of the long-time dynamics, and their relation with the spectral properties of the Lindbladian~\cite{sbmd-23,glp-24,Marche-24}.

\paragraph{\textbf{Acknowledgements} ---}

We are indebted to M.~Fagotti for enlightening discussions on this and related projects.
We are grateful to M.~C.~Ba\~{n}uls, B.~Buca, H.~Katsura, M.~Schir\`o and L.~Zadnik for several discussions.
L.M.~thanks also R.~Fazio and J.~Keeling for uncountable discussions on many-body open quantum systems.
L.C. acknowledges support from ERC Starting Grant 805252 LoCoMacro. 
This work has benefited from a State grant
as part of France 2030 (QuanTEdu-France), bearing the
reference ANR-22-CMAS-0001 (G.M.), and is part of HQI
(www.hqi.fr) initiative, supported by France 2030 under the French National Research Agency award number
ANR-22-PNCQ-0002 (L.M.). This work is supported
by the ANR project LOQUST ANR-23-CE47-0006-02 (L.M. and L.C.). This work was carried out in the framework of the joint Ph.D. program between the CNRS and the University of Tokyo (A.M.). This work was supported by the Swiss National Science Foundation under Division II grant 200020-219400 (L.G.).

\begin{center}
\begin{large}
\textbf{End Matter}
\end{large}
\end{center}

\paragraph{\textbf{Commutant algebra} ---}

We discuss the properties of the commutant algebra (reviewed in~\cite{mm-22,mm-23}) associated with the model in Eq.~\eqref{eq:model} of the main text. Such analysis is particularly useful for open quantum systems, as shown in Ref.~\cite{LiSalaPollmann_2023}, since it allows us to identify in general a set of (both local and non-local) symmetries associated with the terms appearing in the Lindbladian for a generic choice of the couplings. 
For instance, given the Hamiltonian $H = \sum_j h_j$ and the local dissipators $L_j$, one defines the commutant algebra $\mathcal{C}$
\be\label{eq:C_def}
\mathcal{C} := \{\mathcal{O} | \  [\mathcal{O},h_j] = [\mathcal{O},L_j] = [\mathcal{O},L^\dagger_j]= 0 \ \forall j \}.
\ee
We observe that the dissipators of~\eqref{eq:model} are hermitian, $L^\dagger_j = L_j$, and both $\mathds{1}$ and $\ketbra{\Uparrow}{\Uparrow}$ belong to $\mathcal{C}$. Specifically, one can show that $\mathcal{C} = \text{Span}\{\mathds{1}, \ketbra{\Uparrow}{\Uparrow}\}$, meaning that these operators generate the entire commutant algebra. We sketch the proof below:

\begin{itemize}
\item Since we are interested in $g,J\neq 0$ being generic, we will focus on operators that commute with both the hopping and the East-West term in the Hamiltonian. 
\item We identify the operators that commute with $\sigma^z_j,\sigma^z_{j+1}, (1-\sigma^z_j)\sigma_{j+1}^x,  (1-\sigma^z_{j+1})\sigma_{j}^x$ at a given position $j$. One can check that these also commute with $\sigma^+_j\sigma^-_{j+1}+\sigma^-_j\sigma^+_{j+1}$.
\item Finally, one intersects progressively the spaces of operators introduced above over the site index $j$. The result of this procedure, which can be obtained using induction over the length of the chain, gives precisely $\text{Span}\{\mathds{1}, \ketbra{\Uparrow}{\Uparrow}\}$.
\end{itemize}
As a final remark, we point out that, for specific choices of the couplings, as for example $g=0$ (where $\sum_j \sigma^z$ is conserved), additional symmetries can be present and the commutant algebra becomes larger: we consider that as a fine-tuning. In particular, since $\mathcal{C}\subseteq \ker{\mathcal{L}}$ (from the definition~\eqref{eq:C_def}) and a double degeneracy of the zero-eigenvalue is observed at $g,J\neq 0$ (that is, $\text{ker} \mathcal{L} = \text{Span}\{\mathds{1}, \ketbra{\Uparrow}{\Uparrow}\}$), one safely concludes that $\mathcal{C} = \text{Span}\{\mathds{1}, \ketbra{\Uparrow}{\Uparrow}\}$. Additional details are given in the SM~\cite{ref_SM}.

\paragraph{\textbf{Spectral properties of the Lindbladian} ---}

\begin{figure}[t]
 \includegraphics[width=\columnwidth]{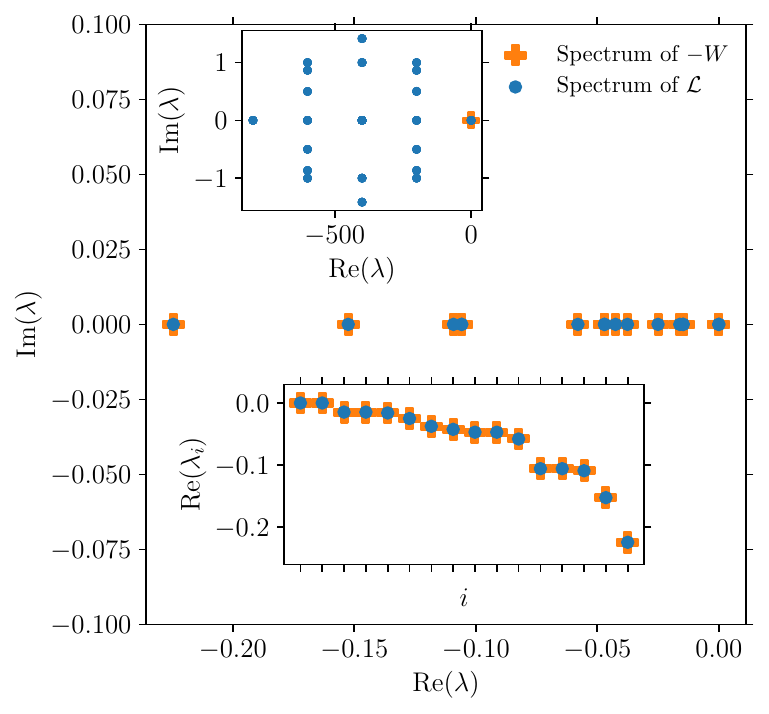}
 \caption{Comparison between the slow decaying modes of the Lindbladian $\mathcal{L}$ and the spectrum of the effective Hamiltonian~$-W$. The parameters are $[J,g,\gamma,L] = [0.5,0.5,100,4]$. In the top inset, the full spectrum of $\mathcal{L}$ is represented. In the bottom inset, we show the real part of the slow decaying modes labeled by the index $i$.}
 \label{fig:compLvsW}
\end{figure}

Here, we discuss the spectral properties of the Lindbladian in Eq.~\eqref{eq:model}. We first, summarize the main steps to obtain the effective Markov generator $W$ in Eq.~\eqref{eq:W}, in the limit of small $g/\gamma,J/\gamma$. The method is based on second-order perturbation theory on the Lindbladian, and it has been employed in~\cite{cb-13}:
\begin{itemize}
\item One first identifies the spectrum of the dissipator, found for $g,J=0$ in Eq.~\eqref{eq:model}. An explicit calculation shows that matrix elements of the form $\ketbra{\boldsymbol{\sigma}}{\boldsymbol{\sigma}'}$, with $\boldsymbol{\sigma},\boldsymbol{\sigma}'$ spin configurations in the $z$-direction, are eigenvectors and their eigenvalue is proportional to the number of mismatches between $\boldsymbol{\sigma}$ and $\boldsymbol{\sigma}'$. In particular, in the absence of perturbation, the manifold of stationary states, associated with $\lambda=0$, is generated by the diagonal elements $\ketbra{\boldsymbol{\sigma}}{\boldsymbol{\sigma}}$.
\item For small $g,J\neq 0$, the large degeneracy of the Lindbladian is lifted and, at the lowest order, the effective couplings between unperturbed eigenvectors are given by virtual transitions onto unperturbed eigenspaces.
\item Specifically, from the effective couplings between the diagonal elements $\ketbra{\boldsymbol{\sigma}}{\boldsymbol{\sigma}}$, one can identify an effective Lindbladian which describes the low spectrum of $\mathcal{L}$ (and having the same matrix elements of $-W$). This allows us to reduce the complexity of diagonalizing a $4^L\times 4^L$ matrix, associated with $\mathcal{L}$, to the diagonalization of a $2^L\times 2^L$ matrix, corresponding to $W$, as far as the low-spectrum is concerned.
\end{itemize}
The explicit calculations to obtain the expression of $W$ are straightforward but lengthy, and they are reported in the SM~\cite{ref_SM}. As a byproduct, we perform a numerical check and we compare the spectrum of $\mathcal{L}$ and that of $-W$ obtained via exact diagonalization using the open-source Python package QuSpin~\cite{quspin_1,quspin_2}; in Fig.~\ref{fig:compLvsW} we show the results for $L=4$, with $[g,J,\gamma] = [0.5,0.5,100]$.

\begin{figure}[t]
 \includegraphics[width=\columnwidth]{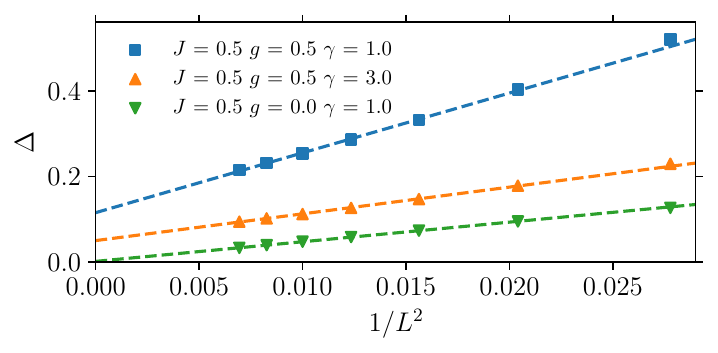}
 \caption{Lindbladian gap $\Delta$ as function of $1/L^2$ for $L \in \{6,\ldots,12 \}$ and periodic boundary conditions.}
 \label{fig:gap_scaling_pbc}
\end{figure}

We now discuss the properties of the lowest energy states of the Markov generator above the ground state manifold. First, we observe that $W\geq 0$ and has $\ket{\Uparrow}$ and $\ket{\Rightarrow}$, with every spin aligned along the $z$- and $x$- axis respectively, as exact eigenstates with vanishing energy (corresponding with the exceptional stationary state and the infinite temperature state, belonging to the kernel of the Lindbladian). Being this operator gapped, as shown numerically in Fig.~\ref{fig:liouvillian_gap}, one expects that its first excited states correspond to domain-wall excitations, interpolating between the two ground states, at given momenta (a property that we found analytically, using the techniques introduced in Ref.~\cite{Tasaki-20, Knabe-88}, for the simplified model $\tilde{W}$ introduced in the main text): while this is formally correct for infinite systems, carefulness is needed for finite systems, and the choice of boundary conditions plays a role. For instance, while a single domain wall can be present with open boundary conditions, only an even number of them can be hosted in the periodic chain: this is analogous to the properties of the spectrum of the quantum Ising chain \cite{Sachdev-99}. As a consequence, the origin of the spectral gap is traced back to a $1-$ or $2-$domain wall excitation with the lowest possible momenta (allowed from the quantization induced by the boundary conditions), for the open/periodic chain respectively. While the previous discussion refers to the Markov generator, we expect that similar properties holds for the Lindbladian. In particular, a discrepancy between the spectral gap with the two aforementioned boundary conditions is explicitly observed: in Fig.~\ref{fig:gap_scaling_pbc} we show the results with periodic boundary conditions, that have to be compared with those in Fig.~\ref{fig:liouvillian_gap}

%%%%FIG
%%%%%

%%%%FIG
%%%%%

\paragraph{\textbf{Details on the tensor-network numerical computations} ---}
Here, we give some details on the implementation of the tensor-network numerical simulations. We vectorize the density matrix $\rho$, as done in Refs.~\cite{Verstraete2004,Cui2015}. In this representation, the Lindbladian~$\mathcal{L}$ becomes a linear operator. We represent the vectorized $\rho$ as a matrix-product-state and $\mathcal{L}$ as a matrix-product-operator. The time-evolution is performed using the Time-Dependent Variational Principle (TDVP)~\cite{Schollwock2019, Haegeman2011, Haegeman2016}. In this way, we are able to simulate chains of $L=40$ sites up to time $t=30$ for $[J,g,\gamma]=[0.5,0.5,1]$. The bond-dimension, the number of sweeps and the time-step used in all this work are $\chi=200$, $n_{\rm sweeps} = 5$ and $\delta t= 0.1$ respectively.

\paragraph{\textbf{On the exponential loss of coherence between $\mathcal S_{\Uparrow}$ and $\mathcal T_{\Uparrow}$} ---}

We discuss here the exponential loss of coherence between $\ket{\Uparrow}$ and the states orthogonal to it, expressed by the quantity $\eta(t)$ in Eq.~\eqref{eq:eta_t}. To study its behavior, we consider the evolution of $\bra{\boldsymbol{\sigma}}\rho(t)\ket{\Uparrow}$, with $\ket{\boldsymbol{\sigma}} \neq \ket{\Uparrow}$. A simple calculation, starting from Eq.~\eqref{eq:model} gives
\be\begin{split}
&\frac{d}{dt}\bra{\boldsymbol{\sigma}}\rho(t)\ket{\Uparrow} = \\
&\bra{\boldsymbol{\sigma}}-i[H,\rho(t)]+\gamma \sum_j (\sigma^z_j\rho(t)\sigma^z_j(t)-\rho(t)) \ket{\Uparrow}.
\end{split}
\ee
Using $H\ket{\Uparrow} =0$ and $\sigma^z_j\ket{\Uparrow} = \ket{\Uparrow}$, we express
\be
\frac{d}{dt}\bra{\boldsymbol{\sigma}}\rho(t)\ket{\Uparrow} = -\sum_{\ket{\boldsymbol{\sigma'}}\neq \ket{\Uparrow}} \bra{\boldsymbol{\sigma}}K\ket{\boldsymbol{\sigma'}}\bra{\boldsymbol{\sigma'}}\rho(t)\ket{\Uparrow} 
\ee
with $ K = iH+\gamma \sum_j (1-\sigma^z_j)$. In other words, the evolution of $\bra{\boldsymbol{\sigma}}\rho(t)\ket{\Uparrow}$, seen as a column vector with entries parametrized by $\boldsymbol{\sigma}$, is ruled by the matrix $K$. We observe that $K+K^\dagger = 2\gamma \sum_j (1-\sigma^z_j)$, and therefore the eigenvalues of $K+K^\dagger$ are $\geq 4\gamma$ in the subspace orthogonal to $\ket{\Uparrow}$. As a consequence, the entries of $\bra{\boldsymbol{\sigma}}\rho(t)\ket{\Uparrow}$ are expected to decay exponentially with a rate proportional to $\gamma$. A quantitative estimate can be provided for the quantifier $\eta(t)$ and, after straightforward algebra, we obtain
\be\begin{split}
-\frac{d}{dt}\eta(t) &= \bra{\Uparrow}\rho(t)(K+K^\dagger)\rho(t)\ket{\Uparrow}  \geq 4\gamma\eta(t)
\end{split}
\ee
which implies $\eta(t)\leq e^{-4\gamma t}\eta(0)$.

\paragraph{\textbf{Finite-size matrix elements of $\tilde{W}$ and non-hermitian skin effect} ---}

We discuss the matrix elements entering Eq.~\eqref{eq:diff_ss1}. We first remark that the space of 1-domain wall configurations is closed under time evolution induced by $\tilde{W}$; thus, it is sufficient to project $\tilde{W}$ onto the associated sector, of dimension $L$, and diagonalize the resulting $L\times L$ matrix. Its diagonalization can be performed efficiently, as it reduces to a single-particle problem with open boundary conditions.

A similar problem appears in the Hatano-Nelson model \cite{hn-96} which is a non-hermitian free fermionic chain; there, the imbalance between the left and right hopping rate is known to result in the localization of the eigenfunctions, a mechanism known as \textit{non-hermitian skin effect} \cite{Martinez-18,yw-18}. The same phenomenology is found in our model and, in particular, we have checked that the right/left eigenfunctions are localized around the right/left boundary respectively. This is sufficient to explain the exponential growth of the matrix elements in Eq.~\eqref{eq:diff_ss1}, as we explain below.

Let us consider a simple ansatz $(x|n) \propto e^{-k(L-x)}$, $(\tilde{n}|x) \propto e^{-kx}$ with $k$ a complex number (depending on $n$) satisfying $\text{Re}(k) >0$. We focus on an observable with support on the right of $x_0$, choosing for simplicity $\mathcal{O} = |x)(x|$, corresponding to the probability distribution of the interface at position $x$. We estimate
\be\label{eq:ma_elem}
(\mathds{1}|\mathcal{O}|n)(\tilde{n}|x_0) \simeq \frac{e^{-k(L-x)}e^{-k x_0}}{\sum_{x'} e^{-k(L-x')}e^{-kx'}} \sim \frac{e^{k(x-x_0)}}{L},
\ee
where the denominator comes from the normalization condition $(\tilde{n}|n)=1$. As a consequence, if $x-x_0 \sim L$ the absolute value of Eq. \eqref{eq:ma_elem} is exponentially large in $L$.

Finally, we point out that while the sum in Eq.~\eqref{eq:diff_ss} converges in the thermodynamic limit, the corresponding sum of the absolute values in Eq. ~\eqref{eq:diff_ss1} diverges exponentially in $L$. The origin of this discrepancy can be traced back to the phases of the matrix element \eqref{eq:ma_elem}, which are highly oscillating as functions of $n$.

\bibliography{bibliography}

%%%%%%%%%%%%%%%%%%%%%%%%%%%%%%%%%%%%%%%%%%%%
\onecolumngrid
\break
\begin{center}
    {\large \bf Supplemental Material: \\
            ``Exceptional stationary state in a  dephasing many-body open quantum system''
}
\end{center}

\appendix

\onecolumngrid

\section{Commutant algebra}

In this section, we discuss the properties of the commutant algebra (reviewed in \cite{mm-22,mm-23}) associated with the model in Eq. \eqref{eq:model}. We will prove that the commutant algebra of \eqref{eq:model} is generated by the two stationary states $\mathds{1}$ and $\ketbra{\Uparrow}{\Uparrow}$.

Before entering the calculations, we briefly review some basic definitions and properties of algebras and their commutants. Given a set of operators $\{\mathcal{O}_j\}$, we denote the ($C^*$-)algebra they generate by $\text{Alg}\l \{\mathcal{O}_j\}\r$: this is a vector space of operators that is closed under products and the adjoint operation $^{\dagger}$ and it contains the identity $\mathds{1}$. For any algebra of operators $\mathcal{A}$, we introduce its commutant
\be\label{eq:commutant_def}
\mathcal{C}[\mathcal{A}] \equiv \{c | \ [c,a]=0 \ \forall a \in \mathcal{A}\}.
\ee
Given two algebras $\mathcal{A}_1,\mathcal{A}_2$, we consider  $\mathcal{A}_1\cdot \mathcal{A}_2 \equiv \text{Alg}\l \{\mathcal{A}_1,\mathcal{A}_2\}\r$ the algebra generated by their products: one can show from the definition \eqref{eq:commutant_def} that
\be\label{eq:commutant_prop}
\mathcal{C}[\mathcal{A}_1\cdot \mathcal{A}_2] = \mathcal{C}[\mathcal{A}_1] \bigcap \mathcal{C}[\mathcal{A}_2].
\ee
In other words, the commutant of $\mathcal{A}_1\cdot \mathcal{A}_2$ contains the operators that commute with both $\mathcal{A}_1$ and $\mathcal{A}_2$. Lastly, we recall that, in the context of open quantum systems, it is convenient to associate the commutant algebra with the local terms of Lindbladians, that is
\be\label{eq:comm_algebra}
\mathcal{C} \equiv \mathcal{C}[\text{Alg}\{h_j,L_j\}_j].
\ee
Here, $h_j$ is the Hamiltonian density, satisfying $H = \sum_j h_j$, and $L_j$ are the jump operators entering the dissipative term of the Lindbladian. In particular, from \eqref{eq:model} and \eqref{eq:comm_algebra}, one gets $\mathcal{L}[\mathcal{C}] = 0$, meaning that $\mathcal{C} \subseteq \text{Ker}(\mathcal{L})$ and the elements of the commutant algebra generate stationary states. 

We now focus on the model \eqref{eq:model} and, with a direct calculation, we show directly
\be\label{eq:Commutant_L_dis}
\text{Span}\{\mathds{1},\ketbra{\Uparrow}{\Uparrow}\} \subseteq \mathcal{C},
\ee
since both the stationary state and the scar are annihilated by each term of the Lindbladian. In the remaining part of this section, we will prove that equality holds in Eq. \eqref{eq:Commutant_L_dis}. To do so, we follow this strategy:
\begin{itemize}
\item We identify the commutant algebra associated with operators inserted on a given site $j=j_0$. Specifically, since the Hamiltonian density contains $2$-site operators, the associated operators have support on $j_0$ and $j_0+1$. 
\item We intersect it with the commutant algebra associated with insertions at $j=j_0+ 1$.
\item We progressively reiterate this process, by intersecting commutant algebras at any position $j$.
\end{itemize}
Without loss of generality, we consider $j_0 =1$, restricting the analysis on the operators acting on $j=1,2$ (isomporhic to $\text{End}(\mathbb{C}^2\otimes \mathbb{C}^2)$). Here, straightforward linear algebra shows that
\be
\begin{split}
&\mathcal{C}[\text{Alg}\{\sigma^z \otimes 1, 1\otimes \sigma^z\}] = \text{Span}\{ \ketbra{\uparrow \uparrow}{\uparrow \uparrow}, \ketbra{\uparrow \downarrow}{\uparrow \downarrow},\ketbra{ \downarrow \uparrow}{\downarrow \uparrow }, \ketbra{\downarrow \downarrow}{\downarrow \downarrow}\},\\
&\mathcal{C}[\text{Alg}\{(1-\sigma^z) \otimes \sigma^x\}] = \l\ketbra{\uparrow}{\uparrow} \otimes \text{End}(\mathbb{C}^2)\r \oplus \l\ketbra{\downarrow}{\downarrow} \otimes \text{Span}\{\ketbra{\rightarrow}{\rightarrow},\ketbra{\leftarrow}{\leftarrow}\}\r,\\
&\mathcal{C}[\text{Alg}\{ \sigma^x \otimes (1-\sigma^z) \}] = \l \text{End}(\mathbb{C}^2)\otimes \ketbra{\uparrow}{\uparrow} \r \oplus \l \text{Span}\{\ketbra{\rightarrow}{\rightarrow},\ketbra{\leftarrow}{\leftarrow}\} \otimes \ketbra{\downarrow}{\downarrow} \r.
\end{split}
\ee
We intersect the three spaces above and the result is 
\be\label{eq:alg_comm_12}
\mathcal{C}[\text{Alg}\{\sigma^z \otimes 1, 1\otimes \sigma^z, (1-\sigma^z) \otimes \sigma^x, \sigma^x \otimes (1-\sigma^z)\}] = \text{Span}\{\mathds{1},\ketbra{\uparrow \uparrow}{\uparrow \uparrow}\};
\ee
this is explicitly the commutant \footnote{Here, we did not consider the commutant coming from the hopping term in Eq. \eqref{eq:H}. The reason is that its (large) commutant contains already $\mathds{1}$ and $\ketbra{\Uparrow}{\Uparrow}$: therefore we do not get any additional information from its intersection with \eqref{eq:alg_comm_12}.} for a chain of length $L=2$. We now consider the operators with support on $j=1,2,3$ as elements of $\text{End}\l (\mathbb{C}^{2})^{\otimes 3}\r$, coming from the insertions at $j=1,2$. We embed \footnote{A technical point is needed here. Given $\mathcal{A}_1 \subseteq \text{End}\l\mathcal{H}_1\r$, one can always construct the embedding $\mathcal{A}_1 \otimes \mathds{1}_{\mathcal{H}_2} \subseteq \text{End}\l\mathcal{H}_1\otimes \mathcal{H}_2\r$ for any other Hilbert space $\mathcal{H}_2$. The commutant of the associated embedding is $\mathcal{C}[\mathcal{A}_1 \otimes \mathds{1}_{\mathcal{H}_2}] = \mathcal{C}[\mathcal{A}_1] \otimes \text{End}(\mathcal{H}_2)$.} the commutant \eqref{eq:alg_comm_12} in this space, and similarly the one associated with $j=2$: we intersect them, and, after simple calculations, we obtain $\text{Span}\{\mathds{1},\ketbra{\uparrow \uparrow \uparrow}{\uparrow \uparrow \uparrow}\}$. We repeat this procedure and, by induction, we eventually find the equality in Eq. \eqref{eq:Commutant_L_dis}.

\section{Effective Lindbladian}

Here, we give details regarding the computation of the effective Markov generator $W$ in~\eqref{eq:W}, obtained using perturbation theory. 
It is convenient to consider the doubled Hilbert space $\mathcal{H}\otimes \mathcal{H}^*$, with $\mathcal{H}$ the Hilbert space of the chain and $\mathcal{H}^*$ its dual. In particular, we regard $\rho$ as a vector of $\mathcal{H}\otimes \mathcal{H}^*$ and $\mathcal{L}$ as an operator acting on it, that is $\mathcal{L} \in \text{End}(\mathcal{H}\otimes \mathcal{H}^*)$. In this formalism, we express the Lindbladian in Eq. \eqref{eq:model} as
\be\label{eq:L_doubled}
\mathcal{L} = -i(H\otimes 1^* - 1\otimes H^{*})+\sum_j \big[ L_j \otimes (L^\dagger_j)^*  -\frac{1}{2}(L^\dagger_j L_j) \otimes 1^* -\frac{1}{2}1\otimes (L^\dagger_j L_j)^*\big],
\ee
where the action of $\mathcal{O}_1\otimes \mathcal{O}^{*}_2 \in \text{End}(\mathcal{H}\otimes \mathcal{H}^*)$ on $\rho$ is $(\mathcal{O}_1\otimes \mathcal{O}^{*}_2)\rho = \mathcal{O}_1 \rho \mathcal{O}_2$.
The unperturbed Lindbladian (obtained for $H=0$) is
\be
\mathcal{L}^{(0)} := -2\gamma\sum_j\frac{1-\sigma^z_j\otimes (\sigma^z_j)^*}{2}.
\ee
We identify its eigenspaces
\be
V_n = \text{Span}\{\ketbra{\boldsymbol{\sigma}}{\boldsymbol{\sigma'}} \},
\ee
where $\boldsymbol{\sigma},\boldsymbol{\sigma'}$ are spin-configurations in the $z$-basis which differ by $n$ spins ($n=0,\dots,L$), with eigenvalue $\lambda_n \equiv -2\gamma n$. We denote the associated spectral projector by $\Pi_j$. We apply second-order perturbation, treating perturbatively the commutator $-i[H,\cdot]$. We do so, and we identify an effective Lindbladian $\mathcal{L}_{\text{eff}}$ acting on $V_0$
\be\label{eq:L_eff}\begin{split}
\mathcal{L}_{\text{eff}} \simeq (-i)^2\sum_{n\neq 0} \Pi_0 (H\otimes 1^* - 1\otimes H^{*})\Pi_n \frac{1}{\lambda_0-\lambda_n}\Pi_n (H\otimes 1^* - 1\otimes H^{*}) \Pi_0 =\\
-2 \sum_{n\neq 0} \frac{\Pi_0 (H\otimes 1^*)\Pi_n  (H\otimes 1^*) \Pi_0}{\lambda_0-\lambda_n} + 2 \sum_{n\neq 0} \frac{\Pi_0 (H\otimes 1^*)\Pi_n  (1\otimes H^*) \Pi_0}{\lambda_0-\lambda_n}.
\end{split}
\ee
It is worth observing that, for the specific Hamiltonian $H$ in Eq. \eqref{eq:H}, $(1\otimes H^*),(H\otimes 1^*)$ connects $V_0$ to $V_n$ for $n=1$ (through the East-West term) and $n=2$ (via the hopping term) only. Before entering the calculation of \eqref{eq:L_eff}, which is lengthy, albeit straightforward, we remind that the validity of this approach is justified whenever the spectral gap above $V_0$ persists in the presence of the perturbation ($g,J\neq 0$): this might be an issue in the thermodynamic limit $L\rightarrow \infty$, and only focus in the regime of $g,J$ at fixed $L$.

Following Ref. \cite{cb-13}, we characterize $\mathcal{L}_{\text{eff}}$ through its action over $\ketbra{\boldsymbol{\sigma}}{\boldsymbol{\sigma}}$ and via the equivalence $\mathcal{L}_{\text{eff}} \leftrightarrow -W$. We consider the East-West term in Eq. \eqref{eq:H} first, which generates terms $\propto g^2/\gamma$ in $\mathcal{L}_{\text{eff}}$. We focus on the contributions appearing in the first term in the r.h.s. of \eqref{eq:L_eff}. We compute them obtaining
\begin{itemize}
\item $\left[\sigma^x_j\l\frac{1-\sigma^z_{j+1}}{2}\r \otimes 1^*\right] \Pi_1 \left[\sigma^x_j\l\frac{1-\sigma^z_{j+1}}{2}\r \otimes 1^*\right]\cdot \ketbra{\boldsymbol{\sigma}}{\boldsymbol{\sigma}} = \left[ \l\frac{1-\sigma^z_{j+1}}{2}\r \otimes 1^* \right] \cdot \ketbra{\boldsymbol{\sigma}}{\boldsymbol{\sigma}}$. A similar term from the exchange $j \leftrightarrow j+1$ appears. 
These generate a term $4\frac{(2g)^2}{2\gamma}\sum_j \frac{1-\sigma^z_{j+1}}{2}$ in $W$.
\item $\left[\sigma^x_j\l\frac{1-\sigma^z_{j+1}}{2}\r \otimes 1^*\right] \Pi_1 \left[\sigma^x_j\l\frac{1-\sigma^z_{j-1}}{2}\r \otimes 1^*\right] \cdot \ketbra{\boldsymbol{\sigma}}{\boldsymbol{\sigma}}= \left[\l \frac{1-\sigma^z_j}{2}\r \l \frac{1-\sigma^z_{j+1}}{2}\r \otimes 1^*\right] \cdot \ketbra{\boldsymbol{\sigma}}{\boldsymbol{\sigma}}$. A similar term with $j+1\leftrightarrow j-1,j\leftrightarrow j$ gives the same contribution; finally, we obtain a term $4\frac{(2g)^2}{2\gamma} \sum_j \frac{1-\sigma^z_{j-1}}{2}\frac{1-\sigma^z_{j+1}}{2}$ for $W$.
\end{itemize}
In the second term in Eq. \eqref{eq:L_eff}, these contributions appear:
\begin{itemize}
\item $\left[\sigma^x_j\l\frac{1-\sigma^z_{j+1}}{2}\r \otimes 1^*\right] \Pi_1 \left[1\otimes\l\sigma^x_j\l\frac{1-\sigma^z_{j+1}}{2}\r\r^* \right]\cdot \ketbra{\boldsymbol{\sigma}}{\boldsymbol{\sigma}} = \left[ \l\frac{1-\sigma^z_{j+1}}{2}\r\sigma^x_j \otimes (\sigma^x_j)^* \right] \cdot \ketbra{\boldsymbol{\sigma}}{\boldsymbol{\sigma}}$. A similar term from the exchange $j \leftrightarrow j+1$ appears. 
These generate a term $-2\frac{(2g)^2}{2\gamma}\sum_j \frac{1-\sigma^z_{j}}{2}(\sigma^x_{j+1}+\sigma^x_{j-1})$ in $W$.

\item $\left[\sigma^x_j\l\frac{1-\sigma^z_{j+1}}{2}\r \otimes 1^*\right] \Pi_1 \left[1\otimes\l\sigma^x_j\l\frac{1-\sigma^z_{j-1}}{2}\r\r^* \right]\cdot \ketbra{\boldsymbol{\sigma}}{\boldsymbol{\sigma}} = \left[ \l\frac{1-\sigma^z_{j-1}}{2}\r\l\frac{1-\sigma^z_{j+1}}{2}\r\sigma^x_j \otimes (\sigma^x_j)^* \right] \cdot \ketbra{\boldsymbol{\sigma}}{\boldsymbol{\sigma}}$. A similar term, with $j+1\leftrightarrow j-1,j\leftrightarrow j$ exchanged in the previous formula, gives the same contribution; finally, we obtain the term $-4\frac{(2g)^2}{2\gamma} \sum_j \frac{1-\sigma^z_{j-1}}{2}\sigma^x_j\frac{1-\sigma^z_{j+1}}{2}$ for $W$.
\end{itemize}

For the hopping term in Eq. \eqref{eq:H}, which gives a contribution proportional to $J^2/\gamma$ for $W$, the calculations are reported below. We observe, that $\sigma^{+}_j \sigma^{-}_{j+1}+\sigma^{-}_j \sigma^{+}_{j+1} = \sigma^x_j\sigma^x_{j+1}\l \frac{1-\sigma^z_j\sigma^z_{j+1}}{2}\r$, and it acts in the $z$ basis by flipping the two spins at $j$ and $j+1$ only when they are different. Thus, from the first term in \eqref{eq:L_eff}, we obtain
\begin{itemize}
\item $[(\sigma^{+}_j \sigma^{-}_{j+1}+\sigma^{-}_j \sigma^{+}_{j+1})\otimes 1^*]\Pi_2[(\sigma^{+}_j \sigma^{-}_{j+1}+\sigma^{-}_j \sigma^{+}_{j+1})\otimes 1^*] \cdot \ketbra{\boldsymbol{\sigma}}{\boldsymbol{\sigma}} = \left[\frac{1-\sigma^z_j\sigma^z_{j+1}}{2}\otimes 1^*\right]\cdot \ketbra{\boldsymbol{\sigma}}{\boldsymbol{\sigma}}$. This contributes as $\frac{J^2}{4\gamma}\sum_j \frac{1-\sigma^z_j\sigma^z_{j+1}}{2}$ to $W$.
\end{itemize}
Similarly, from the second term of \eqref{eq:L_eff}:
\begin{itemize}
\item $[(\sigma^{+}_j \sigma^{-}_{j+1}+\sigma^{-}_j \sigma^{+}_{j+1})\otimes 1^*]\Pi_2[1\otimes(\sigma^{+}_j \sigma^{-}_{j+1}+\sigma^{-}_j \sigma^{+}_{j+1})^*] \cdot \ketbra{\boldsymbol{\sigma}}{\boldsymbol{\sigma}} = \left[\frac{1-\sigma^z_j\sigma^z_{j+1}}{2}\sigma^x_j\sigma^x_{j+1}\otimes (\sigma^x_j\sigma^x_{j+1})^*\right]\cdot \ketbra{\boldsymbol{\sigma}}{\boldsymbol{\sigma}}$. This contributes as $-\frac{J^2}{4\gamma}\sum_j \sigma^x_j\sigma^x_{j+1}\frac{1-\sigma^z_j\sigma^z_{j+1}}{2} = -\frac{J^2}{4\gamma}\sum_j (\sigma^x_j\sigma^x_{j+1}+\sigma^y_j\sigma^y_{j+1})$ to $W$.
\end{itemize}
Putting everything together, we arrive at the expression in Eq. \eqref{eq:W}. As a byproduct of the validity of the result, we easily check $\bra{\Rightarrow}W=0$: such a property corresponds, in the Doi-Peliti formalism, to the conservation of the probability. While this is required for a well-defined Markov chain, it is not completely obvious that this property holds consistently through the perturbative analysis.

\section{Properties of the Markov generator $\tilde{W}$}

In this section, we discuss the properties of the operator $\tilde{W}$ defined in the main text.
We write $\tilde{W} = \sum_j \tilde{w}_j$, with $\tilde{w}_j$ acting on $j$ and $j+1$ whose matrix representation (in the basis $\{\ket{\uparrow\uparrow},\ket{\uparrow\downarrow},\ket{\downarrow\uparrow},\ket{\downarrow\downarrow}\}$) is
\be\label{eq:loc_gen_wt}
\tilde{w}_j = \frac{1}{3}\begin{pmatrix} 0 & 0 & 0 & 0 \\ 0 & 2 & -1 & -1 \\ 0 & -1 & 2 & -1 \\ 0 & -1 & -1 & 2\end{pmatrix}.
\ee
In particular, the entries are chosen so that the configuration $\ket{\uparrow\uparrow}$ is stationary, while the other three mix among each other with equal rates. This allows us to identify two sectors, namely $\text{Span}\{\ket{\Uparrow}\}$ and its orthogonal complement: the first is associated with the stationary state $\ket{\Uparrow}$, while the second is associated with
\be
\frac{1}{2^{L}-1}\l \sum_{\boldsymbol{\sigma}}\ket{\boldsymbol{\sigma}} -\ket{\Uparrow}\r = \frac{2^L}{2^{L}-1}\ket{\bullet} - \frac{1}{2^{L}-1}\ket{\Uparrow},
\ee
that is the infinite temperature state $\ket{\bullet}$ with the contribution of $\ket{\Uparrow}$ being subtracted; this subtraction does not play a role in the infinite volume limit $L\rightarrow\infty$, since the discrepancy with $\ket{\bullet}$ is exponentially small in $L$.

No additional stationary states are present at any finite size $L$, and $\text{ker}\tilde{W} = \text{Span}\{\ket{\Uparrow},\ket{\Rightarrow}\}$: to prove it rigorously, we observe that $\tilde{w}_j$ are frustration-free generators, meaning that $\tilde{w}_j\geq 0$. Consequently \cite{Tasaki-20}, the kernel, spanned by the stationary states, is
\be\label{eq:common_ker}
\ker{\tilde{W}} = \bigcap_j \ker{\tilde{w}_j}.
\ee
One can intersect the kernels of $\tilde{w}_j$ explicitly, obtaining the final result via Eq. \eqref{eq:common_ker}. We point out that the same technique can be applied for $W$ in Eq. \eqref{eq:W}, which gives $\text{ker}W = \text{Span}\{\ket{\Uparrow},\ket{\Rightarrow}\}$.

The analysis of the spectral gap of $\tilde{W}$ is less straightforward. Nonetheless, for the specific choice of the generators $\tilde{w}_j$, that are 2-site projectors (satisfying $(\tilde{w}_j)^2 = \tilde{w}_j$) it is possible to use a technique introduced by Knabe in Ref. \cite{Knabe-88}, to prove the existence of a finite spectral gap: we briefly review it here. The method boils down to analyzing the energy spectrum at finite size in open boundary conditions; in particular, we introduce
\be
\tilde{W}_{L} = \sum^{L-1}_{j=1} \tilde{w}_j,
\ee
that acts non-trivially on the first $L$ sites, and, given its spectral gap $E_L$, one can show \cite{Knabe-88}
\be
\underset{L'\rightarrow \infty}{\liminf} E_{L'}\geq \frac{L-1}{L-2}\l E_L - \frac{1}{L-1}\r \ \forall L\geq 3.
\ee
This gives an explicit lower bound to the spectral gap, that can be computed numerically for small system sizes (for example $L=3,4$). In our specific case, we find, by exact diagonalization
\be\label{eq:lower_bound_E}
\frac{L-1}{L-2}\l E_L-\frac{1}{L-1}\r \simeq 0.05719, \qquad L=3,
\ee
that guarantees the finiteness of the spectral gap in the infinite volume limit.

In the rest of this section, we study the (quantum) single-particle problem associated with the propagation of a kink, and we relate it to the spectral gap of $\tilde{W}$. We introduce the states
\be
 \ket{x}_q \equiv  \ket{\rightarrow\dots \rightarrow  \underset{x}{\rightarrow} \uparrow \uparrow \dots \uparrow},
\ee
which differ from $|x)$, introduced in the main text, by a ($x$-dependent) proportionality constant. From Eq. \eqref{eq:loc_gen_wt}, we find that
\be
\tilde{w}_j \ket{\rightarrow \uparrow}_{j,j+1} = -\frac{\sqrt{2}}{3} \ket{\rightarrow \rightarrow}_{j,j+1} -\frac{\sqrt{2}}{3} \ket{\uparrow \uparrow}_{j,j+1}+\ket{\rightarrow \uparrow}_{j,j+1}, \quad \tilde{w}_j\ket{\uparrow \uparrow}_{j,j+1} = \tilde{w}_j\ket{\rightarrow\rightarrow}_{j,j+1}=0,
\ee
implying
\be
\tilde{W}\ket{x}_q = -\frac{\sqrt{2}}{3}\l \ket{x-1}_q+\ket{x+1}_q\r +\ket{x}_q.
\ee
The associated quantum dynamics can be solved exactly as
\be
e^{-i\tilde{W}t}\ket{x=0}_q = \sum_x\psi(x,t)\ket{x}_q, \quad \psi(x,t) = \int^{\pi}_{-\pi}\frac{dk}{2\pi}e^{ikx-iE(k)t}
\ee
and the single-particle dispersion $E(k)$ is
\be
E(k) = 1-\frac{2\sqrt{2}}{3}\cos k>0.
\ee
Interestingly, the lowest energy of the band is $E(k=0) \simeq 0.05719$ which is exactly the lower bound of the gap obtained in \eqref{eq:lower_bound_E} via the Knabe's method: thus, we identify $E(k=0)$ as the spectral gap of the model and the one-kink band as the low energy spectrum.

\section{Other models}

In this section, we show the appearance of the membrane diffusion of the interface in other models that share with the one considered in the main text: 
(i) the presence of an exceptional stationary state ($\ketbra{\Uparrow}{\Uparrow}$ for simplicity), 
(ii) the lack of extensive strong symmetries,
(iii) the fact that the jump operators are Hermitian, and (iv) the existence of a finite Lindbladian gap.
We believe that these features are \textit{sufficient} to observe the phenomenology described in the main text.
Indeed, they ensure the existence of the generic stationary state proportional to the identity and of the scar (others could be present). The initial state obtained by joining these stationary states must relax on time scales diverging with the system size via a membrane process; these dynamics cannot be described by the Lindbladian gap.

\begin{figure}[t]
 \includegraphics[width=0.4\columnwidth]{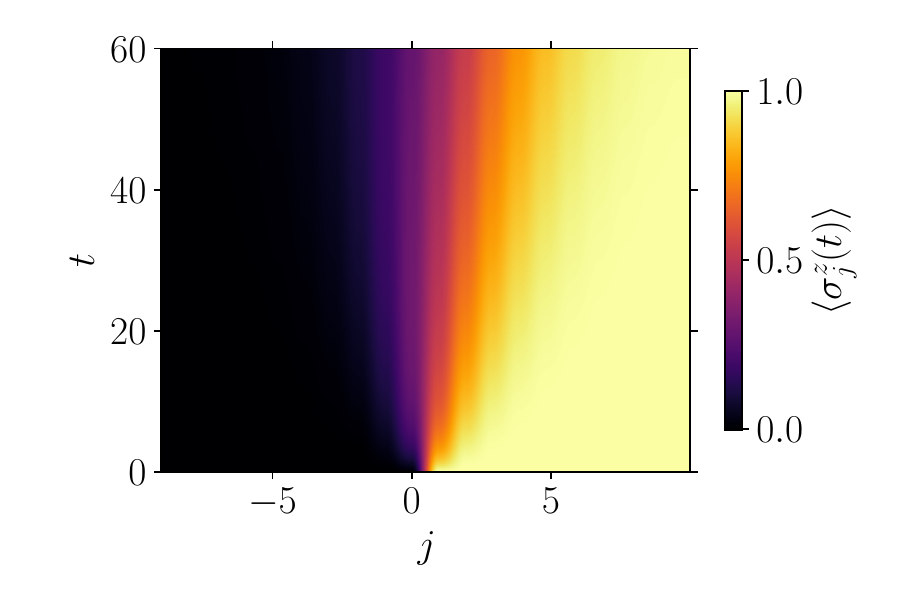}
 \includegraphics[width=0.4\columnwidth]{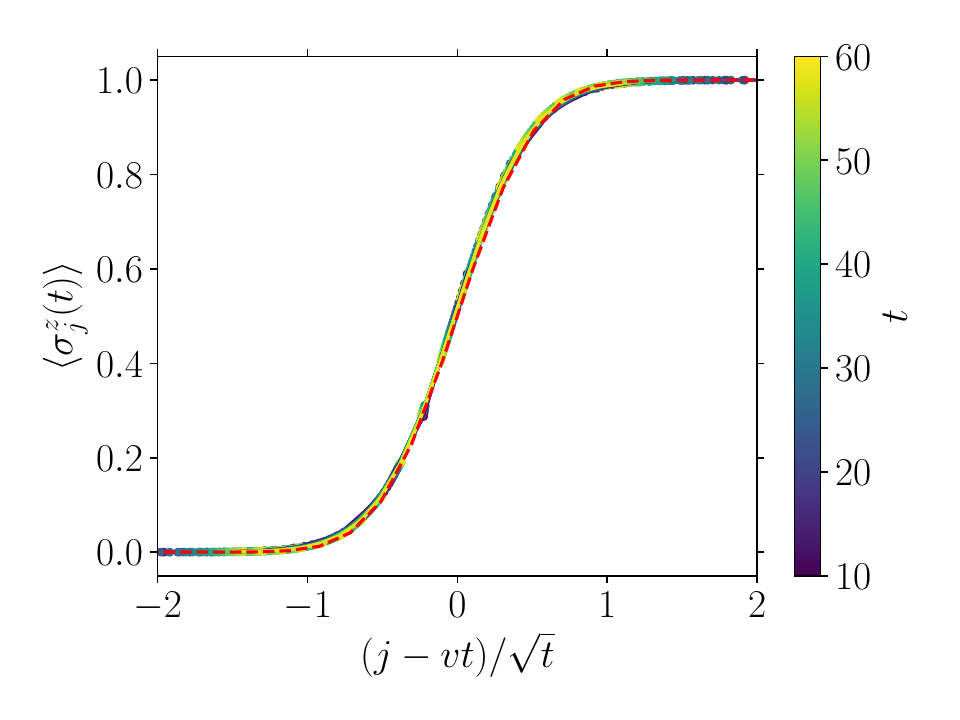}
 \caption{(Left) Profile of the magnetization $\la \sigma^z_j(t)\ra$ for the model in Eq.~\eqref{eq:H1} with parameters $[J,g,\gamma,L] = [0.5,0.5,1,20]$ and open boundary conditions. (Right) Scaling collapse of the magnetization against the prediction~\eqref{Eq:erf:rescaling}, with the parameters $v = 0.023$, $\mathcal{D} = 1.72$ being fitted.}
 \label{fig:xt_PXP_profile}
\end{figure}

We first consider a PXP model with a hopping term and dephasing, whose Hamiltonian and jump operators are
\begin{equation}
\label{eq:H1}
 H_1 =  \sum_j 
J\left[
\sigma^{+}_j \sigma^{-}_{j+1}+H.c.
\right]+
 g \,\pi^z_{j} \sigma^x_{j+1} \pi^z_{j+2};
 \qquad 
 L_j = \sqrt{\gamma} \sigma^z_j .
\end{equation}
We study the time-evolution of $\la \sigma^z_j(t)\ra$ 
for the bipartition protocol of the main text, consisting in putting close by the identity and the scar,
using numerical techniques based on a tensor-network representation of the density matrix and report the results in Fig.~\ref{fig:xt_PXP_profile}.
The left panel shows the space-time dependence of the magnetization profile ahereas the right panel displays a comparison with the scaling function~\eqref{Eq:erf:rescaling} of the main text, finding good agreement. 
In particular, the fitted value of the drift velocity is $v = 0.023$: albeit small, we have carefully checked that the data are not compatible with $v=0$.
This tiny velocity is likely related to the fact that at $J=0$ the initial state  does not evolve, due to the Rydberg constraint; 
non-trivial dynamics emerge only when $J\neq 0$.

\begin{figure}[t]
 \includegraphics[width=0.4\columnwidth]{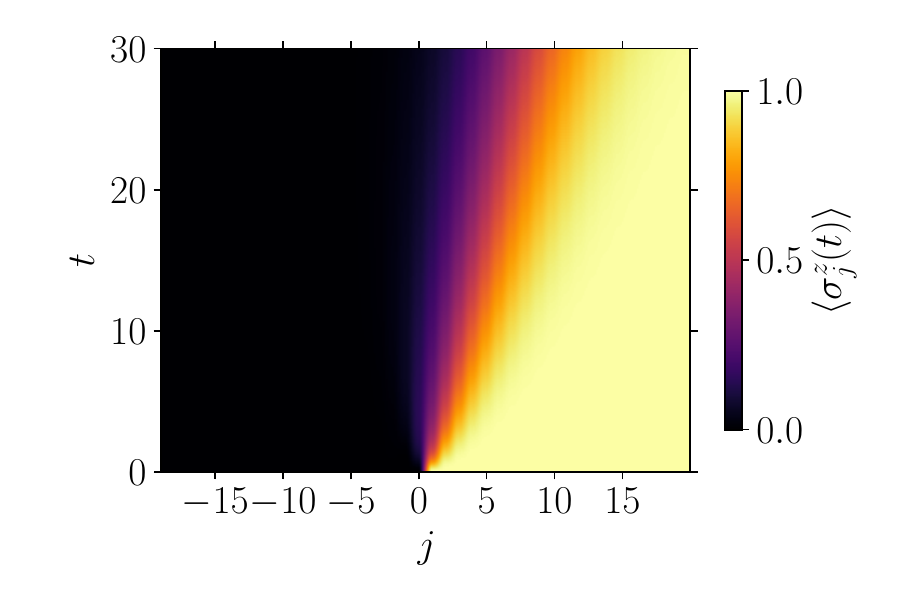}
 \includegraphics[width=0.4\columnwidth]{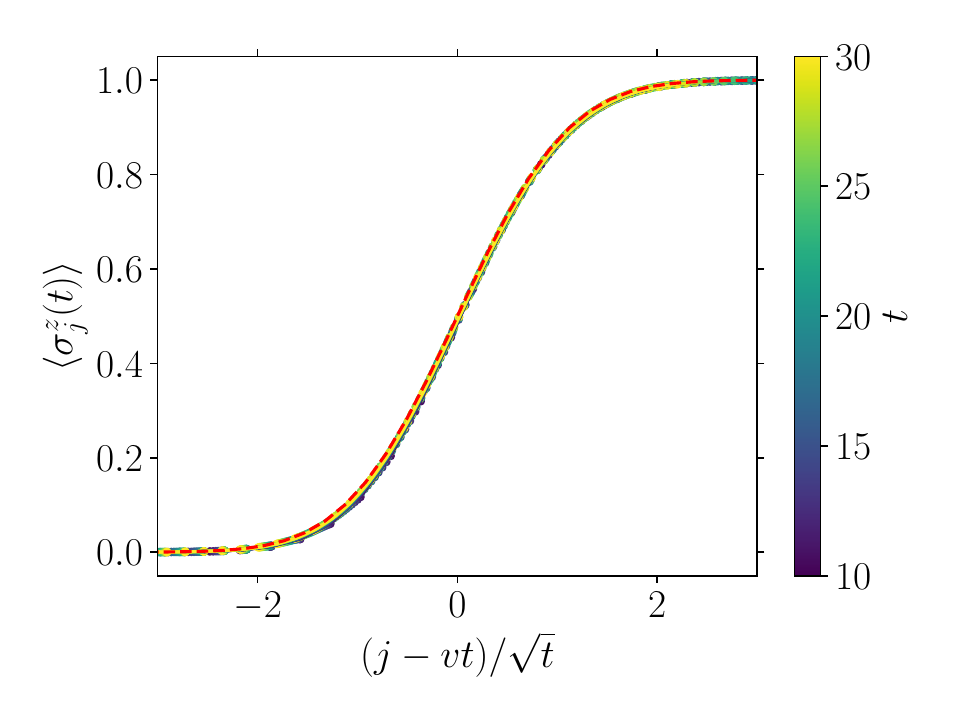}
 \caption{(Left) Profile of the magnetization $\la \sigma^z_j(t)\ra$ for the model in Eq. \eqref{eq:H2} with parameters $[J,g,\gamma,L] = [0.5,0.5,1,40]$ and open boundary conditions. (Right) Scaling collapse of the magnetization against the prediction \eqref{Eq:erf:rescaling}, with the parameters $v = 0.247,  \mathcal{D} = 0.804$ being fitted.}
 \label{fig:xt_XPX_profile}
\end{figure}

We then consider another model, whose Hamiltonian and jump operators are:
\begin{equation}
\label{eq:H2}
 H_2 =  \sum_j 
J\left[
\sigma^{+}_j \sigma^{-}_{j+1}+H.c.
\right]+
 g \, \sigma^x_{j} \pi^z_{j+1} \sigma^x_{j+2},
 \qquad L_j = \sqrt{\gamma} \sigma^z_j.
\end{equation}
We dub the contribution proportional to $g$ as XPX term; the results of our numerical study are shown in Fig.~\ref{fig:xt_XPX_profile}. 
Also in this case, we observe a good collapse with an error function, as predicted by~\eqref{Eq:erf:rescaling}, although the velocity is one order of magnitude larger with respect to the model~\eqref{eq:H1}: for this reason, in order to avoid finite size-effects, we consider a larger system ($L=40$) and smaller time-scales ($t\leq 30$).

\begin{figure}[t]
 \includegraphics[width=0.4\columnwidth]{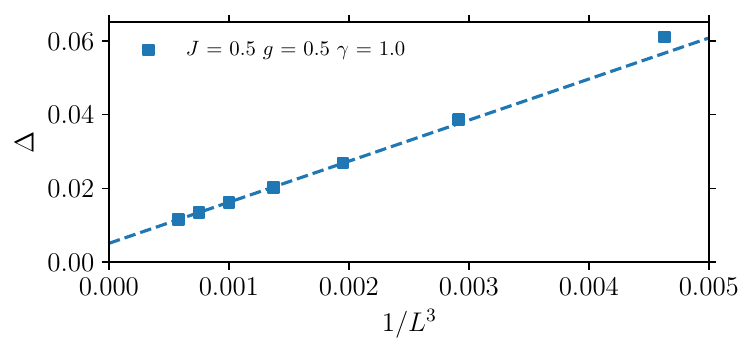}
 \includegraphics[width=0.4\columnwidth]{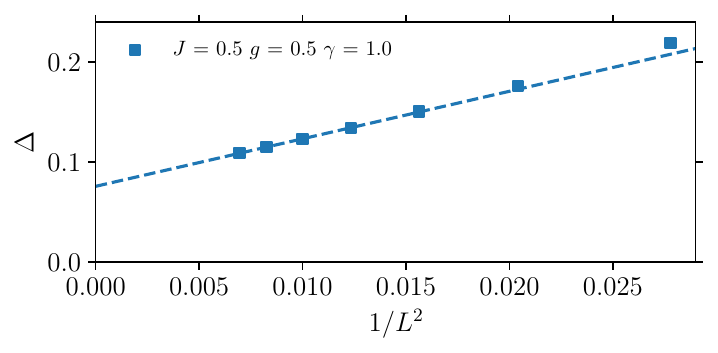}
 \caption{We plot the spectral gap $\Delta$, choosing the parameters $[J,g,\gamma] = [0.5,0.5,1]$ for the PXP and XPX with hopping and dephasing. (Left) Model \eqref{eq:H1}: We fit $\Delta = \Delta_0 + \Delta_1/L^3$, extrapolating $\Delta_0 \simeq 5.1314 \cdot 10^{-3}$ (Right) Model \eqref{eq:H2}: We fit $\Delta = \Delta_0 + \Delta_1/L^2$ and we obtain $\Delta_0 = 7.596\cdot 10^{-2}$.}
 \label{fig:gaps_PXP_XPX}
\end{figure}

Finally, we analyze the Lindbladian gap of the previous models as a function of the system size $L$, showing the results in Fig.~\ref{fig:gaps_PXP_XPX}. 
For the PXP model with hopping~\eqref{eq:H1} we fit $\Delta = \Delta_0 + \Delta_1/L^3$ finding $\Delta_0 \simeq 5.1314 \cdot 10^{-3}$: while the data are not conclusive to exclude gaplessness in the infinite volume limit $L\rightarrow \infty$, we believe that the gap is indeed finite. 
To support our belief, we observe that in the context of the non-Hermitian skin effect it is known that the drift velocity $v$ and the spectral gap $\Delta$ are proportional ($v \sim \Delta$, see Ref.~\cite{Haga_2021}); therefore, the small but non-zero velocity $v$ can be linked to the presence of a small gap. For the XPX model with hopping~\eqref{eq:H2} we propose the scaling $\Delta = \Delta_0 + \Delta_1/L^2$ and our data are consistent with the presence of a finite gap $\Delta_0 = 7.596\cdot 10^{-2}$.

%%%%%%%%%%%%%%%%%%%%%%%%%%%%%%%%%%%%%%%%%%%%
\end{document}